\documentclass[10pt,conference]{IEEEtran} 

\usepackage{amsmath,amsfonts}
\usepackage{xurl}
\usepackage{graphicx}
\usepackage{adjustbox}
\usepackage{tabularx}     
\usepackage{array}
\usepackage{framed}
\usepackage{tcolorbox}
\usepackage{adjustbox}
\usepackage{tikz}
\usepackage{colortbl}
\usetikzlibrary{shapes.geometric}
\usepackage{longtable}
\usepackage[dvipsnames]{xcolor}

\definecolor{AbiOrange}{HTML}{FFD09B}
\definecolor{TimBlue}{HTML}{C4E1F6}
\definecolor{heatblue}{HTML}{1B5FA8}
\definecolor{bargrey}{HTML}{BDBDBD}

\begin{document}

\title{Cognitive Inclusivity in AI Coding Assistants: Evidence from GitHub Copilot}

\author{\IEEEauthorblockN{Grace Kerr, Aarushi Parasrampuria, Faith Culas, Kelly Blincoe}
\IEEEauthorblockA{
\textit{University of Auckland}\\
Auckland, New Zealand \\
{\{}gker710, apar544, fcul804{\}}@aucklanduni.ac.nz, k.blincoe@auckland.ac.nz}}


\maketitle

\begin{abstract}
AI coding assistants are commonly used for daily software development tasks. While prior research has  examined their impact on productivity and code quality, less is understood about how these tools support diverse ways of thinking. Evidence from non-AI software engineering tools suggests that mismatches between tool design and users’ cognitive styles can introduce inclusivity barriers, but it remains unclear whether similar issues arise in AI coding assistants. We investigate how developers with different cognitive styles experience and interact with these tools and whether they exhibit cognitive inclusivity challenges. We conducted a controlled laboratory study with 20 student participants using GitHub Copilot to complete code generation, debugging, and refactoring tasks. Using think-aloud protocols, semi-structured interviews, and inductive thematic analysis, we identify patterns in user interaction. We then apply the GenderMag framework to examine how these patterns differ across cognitive styles. We identify five inclusivity issues concerning how Copilot explains suggestions, supports validation, facilitates learning, exposes its features, and influences the problem-solving process. Our findings suggest that common interaction paradigms in AI coding assistants can privilege certain cognitive styles while introducing additional cognitive effort for others. We derive design implications for more inclusive AI coding assistants. 
\end{abstract}


\section{Introduction}
\label{sec:intro}
AI coding assistants are increasingly integrated into everyday software development workflows, supporting tasks such as code generation, debugging, and refactoring. The 2026 Stack Overflow survey found that 83\% of developers use at least one AI tool in their development process, with AI coding assistants being the most common tool used~\cite{stackoverflow_ai_2026}. Of those using AI coding assistants, 73\% reported using them daily. These tools promise to improve the speed and productivity of software development, and many empirical studies do show substantial productivity benefits~\cite{peng2023impact, cui2026effects,pandey2024transformingsoftwaredevelopmentevaluating,siam2025programming}. However, studies also highlight challenges of using AI coding assistants, including that they can be less effective for complex tasks and large codebases~\cite{pandey2024transformingsoftwaredevelopmentevaluating,pinto2024developer}. 
Much of the early work examining AI coding assistants looked at technical aspects like code correctness, code quality, and security of the outputs~\cite{dakhel2023github, yeticstiren2023evaluating, Ani2024}. However, software engineering is a socio-technical practice~\cite{storey2020software}, and focusing only on the technical outputs does not provide insights into the impacts on teams and individuals.

More recent research examined how developers interact with AI coding assistants, including what factors lead to trust in AI outputs and perceptions of AI use~\cite{brown2024identifying,sergeyuk2025using,vella2026impact}. Emerging evidence suggests that developers do not interact with AI coding assistants in the same way. Experience level shapes how tools are used and who benefits from them~\cite{eyesOnCode2025,shah2025students}. Less experienced developers often ask AI coding assistants to implement entire features rather than decomposing problems on their own~\cite{shah2025students}. Yet, research has not investigated other dimensions of diversity in relation to AI coding assistants use. Given their widespread use, it is important to understand whether these tools can effectively support different development styles or ways of thinking in software development. 

Studies of other tools used in the software development process suggest that inclusivity issues can arise due to differences in cognitive styles across users~\cite{Mendez_GenderMag_2018}. Cognitive styles are the systematic ways in which people process information and approach problem-solving~\cite{kozhevnikov2007cognitive}. Inclusivity issues can often manifest as design mismatches, where tool features inadvertently make it more difficult for users with certain cognitive styles to complete tasks. Although users may still achieve their goals, doing so can require additional work, such as increased cognitive load or the need for additional steps to achieve the same outcome as others~\cite{guizani2022debug}. For example, an evaluation of GitHub identified barriers including lack of visibility, lack of feedback, and information overload, which disproportionately affected users with specific cognitive styles~\cite{santos2023github}. Similarly, cognitive inclusivity issues have been found in other tools like  debuggers and code review tools~\cite{culas2025debuggers,murphy2024gendermag}. 

To date, GenderMag, a framework for identifying inclusivity issues, has been applied to tools that respond predictably to user actions. AI coding assistants generate varied content and offer unprompted suggestions, which may create cognitive-style mismatches that would not appear in prior evaluations. It remains unclear whether the inclusivity issues identified by GenderMag in conventional tools also arise in AI coding assistants, or whether AI coding assistants introduce new forms of inclusivity issues. In this paper, we use the term AI coding assistants to refer to LLM-powered tools~\cite{sergeyuk2025using} embedded in the development environment that provide inline suggestions and chat-based assistance. This includes agent modes where developers remain actively involved in reviewing changes as they are made, but excludes autonomous coding agents that operate independently and present completed changes for later review, such as through pull requests~\cite{li2025teammates}. 

Given the widespread adoption and evolution of AI coding assistants, it is critical to understand their inclusivity issues so that the tools do not disproportionately disadvantage certain users. We address this gap by studying cognitive inclusivity issues in GitHub Copilot, a popular AI coding assistant~\cite{gonccalves2024assessment}, and how they vary across cognitive styles, guided by the following research questions: 
\begin{itemize}
    \item \textbf{RQ1:}  How do student developers experience GitHub Copilot during common software development tasks?
    \item \textbf{RQ2:} How do experiences differ across cognitive styles?
\end{itemize} 

To answer these questions, we conducted a controlled observational study with 20 students using a think-aloud protocol. Participants performed code generation, refactoring, and debugging tasks. Our thematic analysis identified 18 themes on how users experience GitHub Copilot across three areas: Experience, Usage, and Workflow. To examine differences across cognitive styles, we considered the five cognitive facets of the GenderMag framework~\cite{burnett2016gendermag}. We identified five inclusivity issues. 
We discuss these inclusivity issues and design implications for more inclusive AI coding assistants. This study makes the following contributions: 1) Empirical evidence of cognitive inclusivity issues in AI coding assistants, based on a controlled study of student developers using GitHub Copilot; 2) Five inclusivity issues concerning explanation, validation, learning, feature discoverability, and control; and 3) Design implications for more inclusive AI coding assistants.


\section{Background and Related Work}

\subsection{Cognitive Inclusivity in SE Tools}
Previous studies of software, ranging from spreadsheets to basic programming tools, have shown that differences in problem-solving are often associated with gender and cognitive styles~\cite{burnett2016gendermag}. More recent work suggests that such differences extend into professional software engineering workflows, influencing how developers navigate day-to-day tasks~\cite{santos2023github,Mendez_2018,murphy2024gendermag,culas2025debuggers}. Cognitive inclusivity issues have been found in popular tools such as GitHub~\cite{santos2023github,Mendez_2018}, code reviewing tools~\cite{murphy2024gendermag}, and debuggers~\cite{culas2025debuggers,Debugging_2024}. Across these studies, process-oriented learners were often disadvantaged because of unclear workflows (e.g.,  the pull request process in GitHub ~\cite{santos2023github}) and features that were difficult to find without tinkering (e.g., in debuggers~\cite{culas2025debuggers}). Reducing unnecessary cognitive barriers and ensuring cognitive inclusivity in software engineering tools is therefore important for supporting diverse developers. Research also demonstrates that tools can be designed for inclusivity without compromising usability for any group or cognitive style~\cite{vorvoreanu2019gender,murphy2024gendermag}. This highlights the need to find and address cognitive barriers in SE tools, enabling more inclusive development environments.




\subsection{Cognitive Inclusivity and AI coding assistants}
Prior GenderMag studies examined tools whose behaviour was largely deterministic and directly tied to user actions. AI coding assistants differ because they proactively generate suggestions, explanations, and complete solutions, potentially creating new forms of cognitive-style mismatch.
Many studies have found productivity benefits of using AI coding assistants like GitHub Copilot~\cite{peng2023impact, cui2026effects,pandey2024transformingsoftwaredevelopmentevaluating,siam2025programming}, though most treat developers as a homogeneous population. Some studies have examined specific groups like students or novices~\cite{shah2025students,kazemitabaar2023novices,Prather_copilot_2024}. These studies show that user characteristics can influence how AI assistance is used and experienced, but have focused primarily on experience level and educational context rather than other dimensions of diversity, like cognitive style. 

Recent work has begun exploring cognitive inclusivity in AI-assisted development. Anderson et al.~\cite{anderson2026} investigated whether prompting strategies could produce more inclusive explanations. While this work considered cognitive inclusivity, it focused on prompting strategies rather than understanding how developers with different cognitive styles interact with AI coding assistants in practice. As AI tools become increasingly central to software development, it is important to understand whether they support diverse cognitive styles equitably or reinforce existing barriers. We address this gap by examining how students with differing cognitive styles experience GitHub Copilot and by identifying inclusivity issues which unfairly affect certain cognitive styles.

\section{Research Method}
To examine inclusivity issues in AI coding assistants, we conducted a controlled observational study in which participants completed typical software development tasks with GitHub Copilot. We used a think-aloud protocol~\cite{fonteyn1993description} during task completion and conducted post-task interviews to capture participants’ experiences and perceptions. We characterised participants’ cognitive styles using the GenderMag framework~\cite{burnett2016gendermag}. This section describes the study design, participants and recruitment, and data analysis process.

\subsection{Study Design}
Each lab session followed three phases: warmup, task observation, and interview. Each session was conducted individually, with one participant and two researchers present. \par

Each session started with a  \textbf{warmup}, which lasted five minutes. The warmup was designed for the participant to feel relaxed and for the researchers to build rapport with the participant. We collected additional demographic information, including personal pronouns. We provided background information on the tasks, such as the general structure of the tasks.


The next phase was the \textbf{task observation}, which lasted 40 minutes, where participants were asked to complete the three tasks. 
The three tasks were chosen to reflect core software development activities: code generation and refactoring involve producing code, while debugging involves checking and correcting it~\cite{meyer2017worklife,vella2026impact}. Developers commonly use AI assistants across these activities; a recent survey of 481 developers found that 87.3\% of respondents used AI assistants when implementing new features, and around three-quarters used them when refactoring, and two-thirds when bug triaging~\cite{sergeyuk2025using}. 

Participants had access to all Copilot features available at the time of conducting the study: Autocomplete-style suggestions, which appear as grey text in the editor; inline chat, which opens a prompt within the editor; and Copilot Chat, a side panel. Participants could use Copilot Chat in Ask mode, which answers questions without changing code, Edit mode, which proposes changes to files the user selects, or Agent mode, which can edit several files and run terminal commands from a single prompt~\cite{github_modes_blog}. In Edit and Agent modes, changes appeared in the editor for participants to keep or undo. Participants chose which features and modes to use. 

Sessions were conducted using Copilot in Visual Studio Code (VS Code).
VS Code was chosen as the Integrated Development Environment (IDE) as it is the most widely adopted IDE among students at the University. This ensured participants were able to work in a familiar setup, focusing on the tasks and interactions with Copilot rather than the IDE. All three tasks were implemented in the Java programming language and had predefined test cases to enable participants to test their changes and verify their solution. If participants struggled with completing any of the three tasks, up to two hints were offered.
The starter code, instruction scripts, and hint scripts are available in our replication package~\cite{replication_arXiv}. 

Participants were instructed to use the think-aloud protocol by verbalising their thoughts, strategies, and interactions with Copilot throughout the tasks. The researchers demonstrated the protocol through an example. Audio and screen recordings were captured along with observational notes to capture the participant's interactions in detail. 

\textit{First task: } Participants completed a \textbf{code-generation} task by implementing an algorithm from a problem statement and starter code modelled after a  Leetcode-style question~\cite{leetcode_longest_substring}. 
The task assessed how Copilot was utilised during code generation. 

\textit{Second task: } Participants performed a \textbf{code debugging} task involving bugs typical of student data-processing programs, such as incorrect file paths and variable-name mismatches that required diagnosis rather than compilation fixes~\cite{McCall_Novice_Bugs_2019}. Participants fixed bugs until all tests passed. The task assessed how Copilot was used to identify and solve bugs.

\textit{Third task: } Participants \textbf{refactored} existing code, improving the structure and maintainability without changing its behaviour. The code exhibited the Long Method code smell~\cite{fowler1999refactoring}, ensuring a consistent refactoring goal for all participants. This task assessed how participants used Copilot for refactoring.

For all three tasks, the researchers observed whether users accessed GitHub Copilot, which feature and mode they used, how they prompted it, how they interacted with the suggestions provided, whether they accepted or rejected suggestions, and verbal expressions of confidence or hesitation.

The third and final phase was the \textbf{interview}, which lasted 15 minutes. This semi-structured interview was guided by five predefined questions to better understand the participant's experience using GitHub Copilot and identify potential inclusivity issues. We began by asking an open-ended question, '\textit{How did you find the overall experience of using GitHub Copilot?}' We asked additional follow-up questions as needed to clarify responses in greater depth. The full set of predefined questions can be found in our replication package~\cite{replication_arXiv}.

\subsection{Participant Recruitment}
Participants were recruited through two main channels: posters displayed at the University of Auckland and through the University Discord channels. Anyone interested in participating completed a short expression of interest questionnaire, which collected demographic information along with cognitive style information (see Section~\ref{sec:gendermag}). We received 39 expressions of interest. We selected participants purposively to balance across cognitive styles based on their GenderMag facet scores. As more men than women expressed interest, we prioritised respondents who identified as women to improve gender balance, and we favoured respondents with little prior experience of GitHub Copilot to gather early impressions of the tool, though given the popularity of GitHub Copilot this was not always possible. Respondents who were not familiar with Java were excluded, as all tasks were in Java.  In total, we invited 23 individuals, a manageable sample size for qualitative analysis and comparable to other think-aloud studies conducted in software engineering~\cite{culas2025debuggers,SEDHAIN2024101267,RN20}. Of the 23 individuals invited, three did not respond to booking requests while 20 participants booked and completed the full session. Sessions were conducted in July and August 2025.

\subsection{Cognitive Style Classification}
\label{sec:gendermag}
To classify participants' cognitive styles, we used the GenderMag cognitive facet questionnaire~\cite{Hamid2024}. The questionnaire consisted of 14 Likert scale questions. The GenderMag framework uses three personas, Abi, Pat and Tim , to represent differences across five cognitive facets: motivation, information processing style, self-efficacy, risk aversion, and learning style (see Table \ref{tab:GenderMagComparison}). Following the GenderMag framework~\cite{burnett2016gendermag}, we calculated the sample median for each of the five facets and categorised participants' scores as Tim-like (above median), Pat-like (at median), or Abi-like (below median). 




\begin{table*}[t!]
\centering
\caption{GenderMag cognitive facet comparison across personas~\cite{burnett2016gendermag}}
\begin{tabular}{|p{3.5cm}|p{4cm}|p{4cm}|p{4cm}|}
\hline
\cellcolor{bargrey!35}\textbf{Facet} & \cellcolor{heatblue!12}\textbf{Abi} & \cellcolor{heatblue!28}\textbf{Pat} & \cellcolor{heatblue!45}\textbf{Tim} \\ \hline
Motivation & For task completion; prefers familiar methods & Mix of Abi and Tim. & To explore new features \\ \hline
Self-Efficacy & Low self-efficacy; self-blames or gives up. & Medium self-efficacy; persists briefly. & High self-efficacy; blames tool, retries. \\ \hline
Information Processing Style & Reads thoroughly before acting; comprehensive. & Similar to Abi; comprehensive. & Skims; selective. \\ \hline
Learning Style & Structured, process-oriented. & Balanced; cautious experimenter. & Tinkerer; explores quickly. \\ \hline
Risk Attitude & Risk-averse. & Risk-averse. & Risk-taker; enjoys exploration. \\ \hline
\end{tabular}

\label{tab:GenderMagComparison}
\end{table*}

\subsection{Participants}
All participants were students at the University of Auckland and were proficient in Java. 
Of the 20 participants, seven identified as women and 13 as men. Table~\ref{tab:genderMag_facets} shows the distribution of cognitive styles across the five facets. We interpret our findings as characterising developers early in their careers, consistent with recommendations for student-based studies~\cite{falessi2018students}. Two participants reported never using AI coding assistants, 11 used them sometimes, and seven very often. As our analysis examined Copilot interactions across cognitive styles rather than comparing performance between groups, we did not exclude participants with prior Copilot experience.



\begin{table}[b!]
\centering
\caption{Distribution of participants by persona and cognitive facet}
\resizebox{\linewidth}{!}{%
\begin{tabular}{|c|c|c|c|c|}
\hline
\cellcolor{bargrey!35}\textbf{Facet} & \cellcolor{heatblue!12}\textbf{Abi-like} & \cellcolor{heatblue!28}\textbf{Pat-like} & \cellcolor{heatblue!45}\textbf{Tim-like} & \cellcolor{bargrey!35}\textbf{Total} \\
\hline
Learning style & 10 & 0 & 10 & 20 \\
\hline
Information processing style & 10 & 0 & 10 & 20 \\
\hline
Risk style & 8 & 4 & 8 & 20 \\
\hline
Motivation & 9 & 4 & 7 & 20 \\
\hline
Self-efficacy style & 9 & 3 & 8 & 20 \\
\hline
\end{tabular}}

\label{tab:genderMag_facets}
\end{table}

\subsection{Ethical Considerations}
Ethical approval was granted by the University's ethics committee. Participation was voluntary, and participants signed consent forms before the session. All session data were anonymised during transcription by removing personal identifiers and assigning numeric codes to participants (e.g., P1). Participants received a \$20 voucher as a token of appreciation.



\subsection{Study Design Refinement}
Prior to data collection, we conducted a pilot session with one final-year Software Engineering undergraduate student to assess the clarity of the tasks and instructions, timing, and overall study protocol. The participant completed the GenderMag questionnaire prior to the session and successfully completed all phases of the session in approximately 55 minutes. Based on the pilot session, we extended the full session length from 45 to 60 minutes and clarified at the start of each session that participants could use GitHub Copilot however they preferred since this participant expressed uncertainty about how much they were `allowed' to use Copilot. Data from the pilot session were not included in the analysis. 

\subsection{Data Analysis}
Inductive Thematic Analysis~\cite{braun2006using} was used to uncover recurring patterns in participant interactions with GitHub Copilot. Once themes were identified, differences were examined across the cognitive styles using the GenderMag framework.

\subsubsection{Inductive approach using Thematic Analysis}
An Inductive Thematic Analysis method ~\cite{braun2006using,proudfoot2023inductive} was used to enable common patterns and themes to naturally emerge from the session transcripts, without being constrained or influenced by pre-determined categories~\cite{boyatzis1998transforming}. The analysis followed the five phases of thematic analysis: 


\textbf{(i) Data Familiarisation:} Transcripts were automatically generated using Microsoft Teams' live transcription and manually reviewed for accuracy by listening to the audio recordings together with the screen recordings. 
Participants' verbal cues were connected to their interaction with GitHub Copilot. For example, \textit{``I’m about 70\% confident in this. I’m going to keep the changes. $<$Keeps the changes$>$''}. This iterative process developed a strong familiarity with the participant's data, making  subsequent coding and theme development more grounded. Two of the authors transcribed one transcript each and compared them to check that the level of detail was consistent. After agreeing on a shared standard, the remaining 18 transcripts were transcribed independently by two authors. 


\textbf{(ii) Generating the Codeset: } Two authors began by coding the same transcript independently and comparing results to create an initial codeset and align interpretations. This process was repeated for a second transcript, where new codes were merged into the codeset. Two additional transcripts were jointly coded to further align interpretations. The remaining transcripts were then independently coded, followed by two rounds of review across the research team. All coding was completed in NVivo. To preserve contextual meaning, and following Braun and Clarke~\cite{braun2006using}, excerpts could remain uncoded or be coded more than once where relevant. Coding was completed at two levels. Firstly, the semantic level was associated with participants' explicit expression. For example, the statement \textit{``I would not just tell Copilot to refactor for me. I want to see what's actually going on''} was coded as \textbf{`Need to read and understand Copilot's answers before accepting'}. Secondly, the latent level focused on the meaning behind the participant's actions and verbal statements. For example, the statement \textit{``I didn't do anything. Copilot was just like I got you''} was interpreted as trust in Copilot's output and was coded as \textbf{`Trust in Copilot'}. Where meaning was uncertain, audio and screen recordings were reviewed to clarify context before confirming the final codes. Saturation was checked iteratively as the transcripts were coded sequentially, enabling tracking of the emergence of new codes. Thematic saturation was reached by the nineteenth transcript as no new codes emerged thereafter.

\textbf{(iii) Searching for Themes:}
The codes were collated into higher-level themes that captured the overarching recurring ideas of how participants interacted with GitHub Copilot. These themes evolved through several iterative reviews, where related ideas were merged and codes assigned to themes were refined and consolidated. For example, codes that consistently appeared together during the coding process, such as \textit{Information Overload in Copilot Response}, \textit{Long Copilot Response}, and \textit{Suggestion for Copilot Chat to be more succinct}, were grouped under \textbf{Information Presentation}. Where a code was relevant to more than one theme, it was placed under each~\cite{braun2006using}, so the themes are not mutually exclusive. This was done mainly by two authors with iterative discussions with the team.

\textbf{(iv) Reviewing Themes:}
Themes were reviewed to assess how well they reflected the coded data and to ensure they were cohesive and meaningful. They were also reviewed holistically in relation to all data to ensure they accurately represented participants' experiences. This was completed by two authors with iterative discussions with the team.

\textbf{(v) Defining and Naming Themes:}
Each theme was defined and titled to convey its central idea and connection to participants' interactions with the tool. 


\subsubsection{GenderMag Analysis} \label{gendermag-analysis-method}
After generating the themes, we examined differences in participants’ experiences across cognitive styles. Similar to previous studies employing the GenderMag framework, participants generally displayed a mix of cognitive facets rather than aligning with a single persona across all facets~\cite{culas2025debuggers,10.1145/2858036.2858274}. Our analysis therefore examined each facet separately, allowing us to identify facet-specific patterns in how participants experienced GitHub Copilot. Each participant was categorised into personas for each of the five GenderMag facets using the questionnaire described in Section~\ref{sec:gendermag}. We then compared the coded data of Abi-like and Tim-like participants for all theme--facet combinations. For each theme, we read all of the codes from both Abi-like and Tim-like participants looking for differences in experiences or reported preferences. For the first two themes, this was done jointly by two authors. All remaining themes were analysed independently by three authors with iterative discussions between authors. Where we found a difference between the two personas, we recorded an inclusivity issue if a Copilot design feature created barriers or additional effort for one group~\cite{guizani2022debug}. We then grouped these potential inclusivity issues by the associated feature, so that each inclusivity issue relates to a distinct part of Copilot's design. We focus on only Abi-like and Tim-like personas. Comparing the two ends of the spectrum enabled clearer contrasts, and there were too few Pat-like participants to characterise patterns reliably (Table~\ref{tab:genderMag_facets}).

%

\section{Results}

To address our research questions, we first present the common themes of the Copilot user experience (RQ1), followed by how experiences differed across cognitive styles (RQ2).

\subsection{RQ1: GitHub Copilot User Experience}
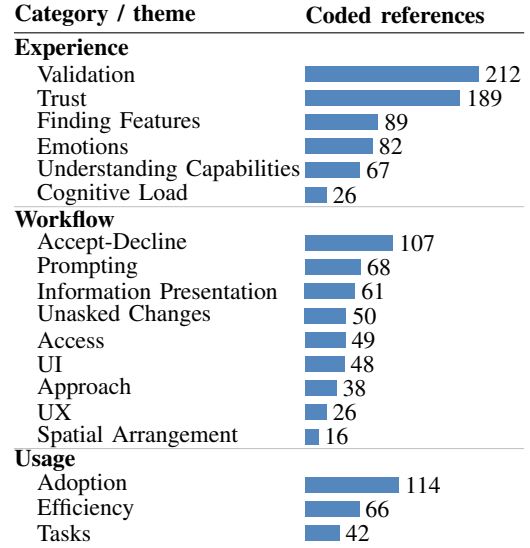
\begin{figure}[t]
  \centering
\def\rowh{0.32}
\def\levelx{0}
\def\namex{0}
\def\indentw{0.3}
\def\barx{3.85}
\def\barmax{2.3}
\def\maxrefs{212}%
 
\definecolor{heatblue}{HTML}{1B5FA8}%
\definecolor{mutedtext}{HTML}{6B6B6B}%
\newcommand{\figfont}{\fontsize{9}{9}\selectfont}%
 
 
\newcommand{\categoryrow}[2]{%
  \node[anchor=west, font=\figfont\bfseries, inner sep=0]
    at (\namex, {-#1*\rowh}) {#2};
}%
 
\newcommand{\grouprow}[2]{%
  \node[anchor=west, font=\figfont\itshape, text=mutedtext, inner sep=0]
    at ({\namex+\indentw}, {-#1*\rowh}) {#2};
}%
 
\newcommand{\themerow}[4]{%
  \pgfmathsetmacro{\rowy}{-#1*\rowh}
  \pgfmathsetmacro{\labelx}{\namex+(#2-1)*\indentw}
  \pgfmathsetmacro{\barlen}{#4/\maxrefs*\barmax}
  \node[anchor=west, font=\figfont, inner sep=0] at (\labelx, \rowy) {#3};
  \fill[heatblue!75!white] (\barx, \rowy-0.1) rectangle ++(\barlen, 0.2);
  \node[anchor=west, font=\figfont, inner sep=0]
    at (\barx+\barlen+0.08, \rowy) {#4};
}%
 
\newcommand{\dividerabove}[1]{%
  \draw[black!25, line width=0.3pt]
    (0, {-#1*\rowh+\rowh/2}) -- (\barx+\barmax+0.6, {-#1*\rowh+\rowh/2});
}%
 
\begin{tikzpicture}
 
\node[anchor=west, font=\figfont\bfseries, inner sep=0] at (\namex, 0.45)
  {Category / theme};
\node[anchor=west, font=\figfont\bfseries, inner sep=0] at (\barx, 0.45)
  {Coded references};
\draw[black, line width=0.4pt] (0, 0.22) -- (\barx+\barmax+0.6, 0.22);
 
\categoryrow{0}{Experience}
\themerow{1}{2}{Validation}{212}
\themerow{2}{2}{Trust}{189}
\themerow{3}{2}{Finding Features}{89}
\themerow{4}{2}{Emotions}{82}
\themerow{5}{2}{Understanding Capabilities}{67}
\themerow{6}{2}{Cognitive Load}{26}
 
\dividerabove{7}
\categoryrow{7}{Workflow}
\themerow{8}{2}{Accept-Decline}{107}
\themerow{9}{2}{Prompting}{68}
\themerow{10}{2}{Information Presentation}{61}
\themerow{11}{2}{Unasked Changes}{50}
\themerow{12}{2}{Access}{49}
\themerow{13}{2}{UI}{48}
\themerow{14}{2}{Approach}{38}
\themerow{15}{2}{UX}{26}
\themerow{16}{2}{Spatial Arrangement}{16}
 
\dividerabove{17}
\categoryrow{17}{Usage}
\themerow{18}{2}{Adoption}{114}
\themerow{19}{2}{Efficiency}{66}
\themerow{20}{2}{Tasks}{42}
 
\end{tikzpicture}
    \caption{Identified themes and the number of coded references for each.} 
  \label{fig:results_overview}
  \end{figure}

Our inductive thematic analysis produced 129 unique initial codes, which resulted in 18 themes across three categories: \textit{Experience}, \textit{Workflow} and \textit{Usage} (see Figure~\ref{fig:results_overview}). While the number of coded references for each theme are provided, frequency should be interpreted cautiously as it may not indicate importance.



\subsubsection{Experience} captures how participants made sense of Copilot and judged its output. 
\textit{Validation} and \textit{Trust} were the most prominent themes in this category. 

\textbf{Validation.}
Participants described varying approaches to assessing and confirming Copilot's output. These differences included their intent to check the output against requirements or their own thought process, the actions they took to validate, and the depth of validation (i.e., how thoroughly these steps were carried out). Actions ranged from immediate acceptance of suggestions to more cautious validation and cross-checking.

\textbf{Trust.} 
Participants revealed varying levels of confidence in Copilot's output. This confidence was reflected in the extent to which they believed that the output was correct, appropriate to the requirements, and aligned with their expectations.

\textbf{Finding Features.} 
Participants varied in how easily they found and navigated Copilot's features, ranging from quickly locating and using them to struggling to discover functionality. 

\textbf{Emotions.}
Participants described a range of emotional responses while interacting with Copilot. These included frustration when suggestions did not align with expectations, hesitation and doubt when deciding whether to rely on Copilot, and feeling overwhelmed by the volume or scope of suggestions. Some participants also showed eagerness to use Copilot, while others felt a sense of reduced engagement or ‘laziness’ associated with relying on Copilot instead of actively working through problems themselves.

\textbf{Understanding Capabilities.} 
Understanding of Copilot's capabilities and features varied. Some participants were confused about the distinction between Autocomplete-style suggestions, inline Copilot, and Copilot Chat, while others expressed uncertainty about how to use Copilot effectively. In contrast, some participants used the range of features without expressing any confusion. 

\textbf{Cognitive Load.}
Participants described increased mental effort arising from lengthy responses, information overload, and effort required to interpret suggestions or explanations.

\subsubsection{Workflow} captures the mechanics of working with Copilot during the tasks. \textit{Accept--Decline}, \textit{Prompting} and \textit{Information Presentation} were the most frequent themes. 

\textbf{Accept--Decline.} 
Participants took different actions and faced varying challenges when accepting, declining, and managing Copilot's suggestions. Some actions included using the `Keep' button and keyboard shortcuts to accept, rolling back changes with undo, and comparing old and new code. Some participants faced difficulty navigating Copilot Chat's suggested changes, particularly when the changes were spread across many lines. 

\textbf{Prompting.} 
Participants varied in how they crafted and refined their prompts to Copilot. These variations included the level of detail provided in prompts, time spent thinking about how to phrase requests, re-prompting in Copilot Chat when initial responses were insufficient, and reluctance to consult Copilot's documentation. 

\textbf{Information Presentation.} 
Participant views on how Copilot presented information ranged from preferring concise, minimal responses to wanting more detailed explanations. Preferences also varied with respect to response structure, readability, and the number of edits suggested at once.

\textbf{Unasked Changes.} 
Participants described instances where Copilot made changes that were not explicitly requested. These included Autocomplete-style suggestions appearing unprompted, Copilot delivering more than asked for, duplicating code, forcing a single solution path, and actions that felt out of place in the context. 

\textbf{Access.} 
Participants' preferences for how they accessed Copilot's functionality varied, including preferences for Copilot Chat, inline Copilot, and Autocomplete-style suggestions. 

\textbf{UI.}
Participants had a range of reactions to the visual design of Copilot's suggestions. These included opinions on the red-green diff layout, the overall look and intuitiveness, and comparisons to other tools such as GitHub and other AI coding assistants.

\textbf{Approach.} 
Some participants were able to incorporate Copilot effectively into their workflow, while others struggled to align Copilot’s outputs with their usual method of working or found it difficult to integrate the assistance provided.

\textbf{UX.} 
Participants described varying experiences using Copilot in terms of ease of use and usability. These ranged from satisfaction with Autocomplete-style suggestions and Copilot Chat to frustration when these disrupted their workflow.

\textbf{Spatial Arrangement.} 
This concerns where Copilot sits within the IDE. Participants faced challenges including insufficient screen space, small button sizes, difficulty viewing code and Copilot outputs simultaneously, and preferences for layouts that enabled clearer side-by-side comparison.

\subsubsection{Usage} captures what participants used Copilot for and the value they saw in it. \textit{Adoption} was the most frequently discussed theme. 

\textbf{Adoption.} 
Participants described varying levels of willingness to use Copilot and intention to continue using it. These ranged from actively incorporating Copilot into their workflow for common tasks to reluctance to rely on it for more complex or unfamiliar problems and resistance to using it overall. 

\textbf{Efficiency.}
Perceived productivity benefits from using Copilot included reductions in time spent completing tasks and satisfaction with the speed of generating solutions, balanced against the effort required to review and validate suggestions.

\textbf{Tasks.}
Participants described using Copilot for a range of purposes during the session, including generating code, debugging errors, and learning through explanations. Usage varied from relying on Copilot for implementation to using it for understanding unfamiliar concepts or code segments.

\smallskip

Based on these findings, we \textbf{answer RQ1}: Participants’ experiences were shaped by multiple factors, including how the participants perceived their interactions (experience), how they integrated Copilot into their tasks (workflow), and the usefulness of Copilot (usage). Validation and trust emerged as central themes.


\subsection{RQ2: Cognitive Style Differences and Inclusivity Issues}

We examined how experiences differed between Abi-like and Tim-like participants for every theme--facet combination. 
Some differences are expected from the persona definitions themselves (Table~\ref{tab:GenderMagComparison}). For example, Abi-like participants reviewing suggestions more thoroughly reflects their comprehensive information processing style. Our analysis therefore focuses on where Copilot's design turns these expected differences into additional effort for some users. Of the 90 theme--facet combinations, 15 showed such a difference, which together revealed five inclusivity issues (Table~\ref{tab:inclusivity_issues}). In the remaining combinations, we found no consistent differences or only differences that were not associated with barriers. 

\begin{table*}[t]
\centering
\caption{Inclusivity issues identified in GitHub Copilot and the themes that informed each issue}
\small
\renewcommand{\arraystretch}{1.15}
\begin{tabularx}{\textwidth}{|
>{\raggedright\arraybackslash}p{2.2cm}|
>{\raggedright\arraybackslash}p{4.45cm}|
>{\raggedright\arraybackslash}p{1.8cm}|
>{\raggedright\arraybackslash}p{5.4cm}|
>{\raggedright\arraybackslash}p{2.1cm}|}
\hline
\rowcolor{bargrey!35}
\textbf{Inclusivity issue} & \textbf{How it appears in Copilot} & \textbf{Disadvantaged group (facet)} & \textbf{Observed impact} & \textbf{Themes} \\ \hline
1. Costly Access to Reasoning & Autocomplete-style suggestions appear in the editor without reasoning. Copilot Chat explains changes in a separate panel, often at length. & Abi-like (information processing style) & Abi-like participants switched between the editor and Chat to find the reasoning behind a suggestion, and found the long responses time-consuming to read. & Validation, Trust, Cognitive Load, Information Presentation \\ \hline
2. Accept-and-Revert Workflow & Suggestions are applied in one action and reversed with `Undo'. & Abi-like (risk style, learning style) & Abi-like participants re-prompted or declined suggestions they were unsure of. Some accepted changes while still unsure whether they would break working code.  & Validation, Emotions, Trust \\ \hline
3. Task Completion Over Learning & Suggestions primarily support task completion rather than broader learning. & Tim-like (motivation) & Tim-like participants who wanted to learn compared suggestions against their own reasoning without support from Copilot. & Validation \\ \hline
4. Hard-to-Discover Features & Features like Chat, Agent mode, and Add Context are mostly found by exploring the interface. & Abi-like (self-efficacy) & Abi-like participants needed help to find features, described their own confusion, and asked for documentation. & Finding Features, Understanding Capabilities \\ \hline
5. Copilot-Led Solutions & Complete solutions are offered as the user works, both in Autocomplete-style suggestions and in Edit and Agent modes. Changes often involve many lines or files. & Abi-like and Tim-like (self-efficacy) & Abi-like participants adopted Copilot's approach in place of their own and described guilt for relying on it. Tim-like participants kept to their own plan but were frustrated when Copilot pushed a single solution. & Validation, Trust, Cognitive Load, Emotions, Approach \\ \hline
\end{tabularx}
\label{tab:inclusivity_issues}
\end{table*}

\subsubsection{Costly Access to Reasoning}

Autocomplete-style suggestions appear in the editor without any reasoning, and Copilot Chat's explanations appear in a separate panel, often as long, detailed responses. People with different information processing styles interacted with these explanations differently. Participants with a Tim-like information processing style (selective) were more likely to judge Copilot suggestions by their outcomes without relying on explanations, while those with an Abi-like information processing style (comprehensive) often wanted to understand the reasoning behind a suggestion before trusting it.

Tim-like participants more often trusted a suggestion once the test cases passed and more often accepted a suggestion first and checked it afterwards. P13 explained, \textit{``If it passes the test cases, then yeah. Pretty confident with it.''} Tim-like participants also tended to read the code rather than Copilot Chat's explanation of it. P5 said, \textit{``I would say that sometimes looking at the code will make more sense than reading it [the explanation from Copilot].''} 

On the other hand, participants with an Abi-like information processing style were more likely to rely on explanations to build trust, and passing test cases were not always enough. P14 said, \textit{``Hey, it says it all passed, but I'm not sure if I trust it.''} Many Abi-like participants reviewed suggestions line by line and compared the chat summary with the code in the editor. P1 noted, \textit{``You'll see on the right in the chat that it explains why it changed, which is really helpful.''} 
However, many Abi-like participants also  commented on the length of these explanations. P16 said, \textit{``Oh my God, that is long. That would take more time than me actually trying to do it myself.''} Some Tim-like participants also found the responses long, but they skimmed them or went straight to the code.
 
As a result, participants with an Abi-like information processing style often switched between the editor and Copilot Chat to find reasoning and expended additional effort to read lengthy responses. Tim-like participants, who processed information more selectively and tended to focus on outcomes, largely avoided this effort.
 
\subsubsection{Accept-and-Revert Workflow}

Copilot's suggestions are applied in a single action, using `Tab' for Autocomplete-style suggestions and `Keep' in Edit or Agent mode, and, if needed, reversed with `Undo'. Users can see the changed code, but not how it will affect the rest of the codebase. Checking whether a suggestion works usually means applying it, running the tests, and undoing it if needed. Participants with a Tim-like risk or learning style were more willing to accept a suggestion before knowing whether it was correct, while Abi-like participants preferred to be certain first. 

Several participants with a Tim-like (tinkering) learning style described accepting and reverting as part of how they built trust. P12 said, \textit{``So the first time, just run it. Just run it. You know, you can always go back.''} Participants with a Tim-like risk style showed little hesitation to use Copilot and accept suggestions. P2 said, \textit{``No, I jump straight into Copilot every time. Yeah, no hesitation there. Love it.''} When participants with a Tim-like risk style did raise concerns, the concerns were often about over-reliance on Copilot in the longer term, not about the suggestion in front of them.
 
Conversely, many participants with an Abi-like (risk-averse) risk style worried about the code in front of them, such as breaking code that already worked. P16 said, \textit{``Sometimes I don't trust
[AI] you know, because it could mess up your entire thing.''} Participants with an Abi-like risk style checked suggestions, re-prompted Copilot when unsure, and in some cases declined suggestions rather than accepting and rolling back. Others accepted a change while still unsure about it. P5 kept a change and noted, \textit{``I'm 30\% not confident because I wasn't sure if it changed the actual functionality.''} P3, who has an Abi-like (process-oriented) learning style, said, \textit{``I still feel like I need to check everything.''}
 
The accept-and-revert workflow suited participants with Tim-like risk and learning styles, who took risks and tinkered, but offered little support to those with Abi-like risk and learning styles for resolving their uncertainty before committing.

\subsubsection{Task Completion Over Learning}
Copilot's default suggestions primarily support completing tasks and provide limited support for broader learning goals, such as exploring alternative approaches. This resulted in differences between participants with different motivations. Participants with a Tim-like motivation (to explore) often saw Copilot's suggestions as a chance to learn or for self-assessment, while those with an Abi-like motivation (for task completion) were more focused on task outcomes. 

Some Tim-like participants validated suggestions against their own reasoning rather than only checking correctness or adherence to task requirements. For example, P20 said, \textit{``This is more so I can learn or test myself if I can understand it before getting a knowledge of how it's refactored I guess,''} and then reviewed the Copilot edits against their own approach. They went on to say \textit{``I would rather use it for repetitive stuff more so I can keep learning while doing it.  So I would use it for things I already know.''}
 
Participants with an Abi-like motivation focused on whether the output met the task requirements. P3 checked whether \textit{``the main functionalities \ldots\ are all good''} by cross-checking against the task description. Some participants with Abi-like motivation were willing to use code they did not fully understand. P1 said, \textit{``I don't fully understand what it's all doing, and I don't want to take the time to actually understand if I know that I have Copilot there to help.''}
 
Participants with Tim-like motivation who wanted to learn largely had to do so through their own effort, as Copilot offered little support for learning-oriented workflows.

\subsubsection{Hard-to-Discover Features}
Copilot's features are mostly found by exploring the interface, with little guidance on where they are or what they do. Participants differed in what they did when they could not find or understand a feature. 

Most participants with Tim-like self-efficacy learned about features by trying things out. P18 said, \textit{``It's kind of just mess around until you find what you want.''} P11 was the only participant with Tim-like self-efficacy who struggled to find Copilot Chat, but they clicked through the interface until they found it, explaining, \textit{``I think the first thing I try would be the side. Right. My second port of call would be accessing it from here. I don't exactly know how to get the chat window. Here, I see ... It's out of the way.''} 
 
More participants with Abi-like self-efficacy had difficulty finding Copilot Chat and some needed help from one of the research observers to find it. Reflecting on finding Copilot's features, P16 said, \textit{``Yes, but only if you knew where it was and what it looked like. But if you didn't then, no.''}  Participants with Abi-like self-efficacy often described their own confusion about what Copilot could do. P1 said, \textit{``I’m just not sure how I would use Copilot to check ''}. P5 suggested better documentation, \textit{``I still don’t know what that means. They should have better documentation, hover messages or an `i' button for information.''}
 
Participants with Tim-like (higher) self-efficacy explored until they found what they needed, while those with Abi-like (lower) self-efficacy often needed help and blamed themselves. 

\subsubsection{Copilot-Led Solutions}
Copilot offers complete solutions that can change many lines or files at once as users work through Autocomplete-style suggestions and in Edit and Agent modes. Participants with Tim-like self-efficacy tended to keep control of their own approach, while those with Abi-like self-efficacy tended to follow Copilot's. 

Participants with Tim-like self-efficacy often kept to their own plan, kept working on parts they found difficult, and were cautious about trusting Copilot. P17 explained, \textit{``I would like to have my own approach first and then see from there. And then if I need more help, I can refine my ideas with it.''} When Copilot pushed a different solution, Tim-like participants were often frustrated. P11 said, \textit{``That's probably the biggest annoyance that it was just trying to force me down one thing.''}
 
Nearly all participants with Abi-like self-efficacy displayed the opposite pattern. They adopted Copilot's suggestions in place of their own ideas, handed over tasks that seemed beyond them, or trusted Copilot because they doubted their own ability. P4 said, \textit{``I wanted to do it by myself at first but it kind of suggested it so I'm just going to go with it.''} Rather than being frustrated with the tool, Abi-like participants described feeling guilty or lazy for relying on it. P20 said, \textit{``Some of, I guess, the hesitation of using it is guilt, that it's doing it for me instead of me doing it myself, so I hesitate to ask it questions if I really feel I should be knowing this.''}

Participants with Abi-like (lower) self-efficacy sometimes accepted these solutions even when they had intended to solve the problem themselves. Those with Tim-like (higher) self-efficacy set the same suggestions aside to follow their own approach.

\smallskip

Based on these findings, we \textbf{answer RQ2}: Abi-like and Tim-like participants differed in their use of explanations, how they accepted suggestions, preferences to learn, how they found features, and their level of reliance. We identified five inclusivity issues. Three mainly disadvantage Abi-like users, who process information comprehensively, prefer certainty before acting, or have lower self-efficacy. One disadvantages Tim-like users who are motivated to learn. The fifth affects both groups, as complete solutions led Abi-like participants to defer and frustrated Tim-like participants.

%

\section{Discussion}
This study identifies five cognitive inclusivity issues in GitHub Copilot.
In this section, we interpret the findings within existing literature, present implications for more inclusive design, and acknowledge threats to validity.

\subsection{Consistency with Prior Studies}
\label{sec:consistency}
Previous studies using the GenderMag framework have shown that Abi-like users tend to experience more difficulties than other personas when interacting with software tools~\cite{Mendez_GenderMag_2018,santos2023github,culas2025debuggers,murphy2024gendermag}. Similarly, three of the five inclusivity issues we identified primarily affect Abi-like users. Issue 4 (hard-to-discover features) mirrors findings from studies of code review tools and debuggers, where lower self-efficacy contributed to difficulties discovering functionality and participants with higher self-efficacy were more likely to overcome these barriers through exploration and tinkering~\cite{culas2025debuggers,murphy2024gendermag}. 

Our results extend prior findings in two important ways. First, AI coding assistants introduce new interaction patterns that are not common in traditional software engineering tools. AI coding assistants generate content proactively, offer unprompted suggestions, and can produce complete solutions. This creates new forms of cognitive-style mismatch that arise from how users interpret, validate, and respond to AI-generated outputs. Several of our inclusivity issues appear to be specific to AI coding assistants. Issue 1 (costly access to reasoning) emerged because suggestions and explanations are separated across different interaction mechanisms. Issue 2 (accept-and-revert workflow) arose because evaluating a suggestion often requires accepting AI-generated changes before assessing consequences. Issue 5 (Copilot-led solutions) stems from the proactive generation of complete solutions, creating tensions around ownership, trust, learning, and control. 


Second, while earlier studies have generally found that software features support Tim-like traits well~\cite{burnett2016gendermag,vorvoreanu2019gender}, Issue 3 (task completion over learning) shows a barrier for users with a Tim-like motivation in which Copilot supported efficient task completion, but little support to use suggestions for learning. 

Taken together, these findings suggest that cognitive inclusivity issues identified in traditional software tools can remain relevant in AI coding assistants, but that AI coding assistants also introduce new inclusivity issues. 
Our findings parallel phenomena reported in the human-AI collaboration literature, including trust calibration~\cite{baltes2026need} and maintenance of human agency~\cite{feng2026junior}. We extend this work by suggesting that cognitive style may help explain why such challenges are experienced differently across individuals. Given the rapid adoption and evolution of these tools, these findings highlight the importance of designing inclusive AI coding assistants.


\subsection{Mechanisms Underlying Inclusivity Issues}

\textbf{Some validation strategies are favoured over others:} Some participants validated suggestions shallowly, while others engaged in deeper validation. This creates an asymmetry in interaction cost: designs that support rapid acceptance workflows align well with shallow validation, but provide limited support for users who require explanation, reasoning, and confirmation before acting. As a result, users with deeper validation strategies incur additional cognitive overhead when interacting with the tool.
This is reinforced by how Copilot applies suggestions. Autocomplete-style suggestions and Edit and Agent modes all prioritise low-friction acceptance (e.g., Tab, Keep, and Undo) without providing inline explanations or showing a suggestion's broader effect before it is accepted.  While this supports fast, outcome-driven workflows, it offers limited support for users who rely on understanding to build trust. Our findings suggest that integrating lightweight explanation and validation support directly into the suggestion workflow is key to enabling more inclusive interaction.

\textbf{Self-efficacy is associated with different responses to AI-generated solutions:} Participants with lower self-efficacy needed help to find Copilot's features and, when Copilot offered complete solutions, often adopted them in place of their own approach and described guilt about relying on it. Participants with higher self-efficacy were more likely to explore and maintain control of their own approach, but were frustrated when Copilot pushed a different solution. Because AI coding assistants can produce complete solutions, how much the tool leads becomes an inclusivity concern in its own right, not only a question of how users validate its output. 

\textbf{Speed is not optimised for all:}  Many interaction patterns that reduced effort for some users, such as immediate suggestions and low-friction acceptance, shifted effort onto users who preferred understanding or certainty before acting. This reinforces the need to account for cognitive diversity when designing future AI coding assistants. Providing easier access to desired information can improve efficiency for more users.


\subsection{Design Implications}
We derive five design implications based on the identified inclusivity issues. These represent potential directions for future tools, not validated solutions. 
Each considers how future AI coding assistants might better support additional cognitive styles without removing the support for current workflows.
 
\subsubsection{Support access to reasoning} Future designs could provide short explanations of why a suggestion was made alongside the suggestion itself, with an option to expand the explanation if desired. 
This brings reasoning closer to the code for users who prefer to understand suggestions before acting, while allowing others to focus on the suggestion itself. Similarly, chat responses could be more concise by default with additional detail on request. This design may reduce the effort required for users who prefer explanations without adding friction for users who prefer outcome-driven workflows.
 
\subsubsection{Support evaluation before acceptance} Future AI coding assistants could support lightweight evaluation before suggestion acceptance by offering an optional view of potential impact of a suggestion on the full codebase. 
This would enable users to judge a suggestion's consequences before committing, while still allowing direct acceptance for those who prefer it.
 
\subsubsection{Support learning-oriented interaction} Future designs could support both task-oriented and learning-oriented workflows, for example by allowing users to toggle between modes. Suggestions could also include an optional `Learn more' link that provides access to learning resources for the concepts or library methods a suggestion uses.
These designs would support users who want to learn, while still enabling quick task completion when learning is not the goal. 
 
\subsubsection{Support feature discoverability} Future AI tools could make features easier to discover by providing lightweight guidance or tutorials for first-time users. For example, the interface could point users to the Copilot Chat window and explain the purpose of features like the mode picker and Add Context button through short descriptions or an information button, as P5 suggested. This design may reduce the need for exploratory trial-and-error and help users who are less likely to explore to still understand the capabilities that are available. 
 
\subsubsection{Support different levels of AI guidance and control} Future designs could offer configurable workflow options, allowing users to choose how much guidance they receive and control they retain. Modes could differ in the amount of code produced at once, number of alternatives presented, solution completeness, and level of information provided.  This would allow users to stay in control when they want to and to reduce feelings of being pushed down a single path. 


These design implications are grounded in observed inclusivity issues but are not validated. Future work should evaluate if they reduce inclusivity barriers without creating new ones. 

\subsection{Beyond GitHub Copilot}
While our study focuses on Copilot, the design mechanisms underlying several of the inclusivity issues are not unique to it. Low-friction acceptance of inline suggestions, single-solution outputs presented without alternatives, and reasoning available only in a separate chat surface are common across current suggestion-based AI coding assistants, including Cursor~\cite{cursor_docs} and JetBrains AI Assistant~\cite{jetbrains_ai}. Thus, the inclusivity issues we found may apply to additional tools. 
We compare our design implications to the current features of AI coding tools available in October 2026 in the latest version of VS Code~\cite{vscode_v1_141}. None of the tools we reviewed explain a suggestion where it appears (Design Implication 1) or enable evaluation before accepting suggestions (Design Implication 2). Some provide partial support for the remaining design directions.
There is limited support for learning-oriented interactions (Design Implication 3). Claude Code offers a built-in learning style that explains its choices and leaves small pieces of code for the user to write~\cite{claude_code_output_styles}. Copilot supports a similar learning mode, but requires users to configure it in copilot-instructions.md~\cite{github_learning_setup}
For Design Implication 4, Copilot provides partial support and now shows experimental tips that introduce features users have not yet tried. But these tips are available only once a chat is open. Copilot also lets users choose how eagerly next edit suggestions appear, providing partial support for Design Implication 5~\cite{vscode_v1_110}, but only eagerness is configurable, not other options like the number of alternatives. 


Overall, many inclusivity issues are likely to remain because current support is often partial,  difficult to discover, or outside primary workflows.

\subsection{Threats to Validity}

\textit{Construct Validity:} The use of a think-aloud protocol may have influenced participant behaviour, as verbalising reasoning and being observed are not natural. However, this is a widely used research method to capture reasoning~\cite{culas2025debuggers,whalley_2023}, and we provided examples during the warmup to help participants become familiar with the protocol. Knowing the study concerned Copilot may have encouraged participants to use it more than they would in their normal workflow. Cognitive styles were classified using a self-reported questionnaire and the group median, so Abi-like and Tim-like are relative to our sample, and participants with scores close to the median may differ little across categories. We excluded Pat-like participants whose scores fell on the median, but they could have inclusion issues that we missed. While GenderMag provided the primary lens, some observed differences may reflect other factors such as prior programming experience or confidence with the task domain rather than cognitive style alone. Coded references may also reflect how much participants verbalised rather than only what they experienced. To mitigate this, we coded participants' actions from the screen recordings alongside their speech and obtained additional perceptions through semi-structured interviews. We identified inclusivity issues by examining the coded data for every theme--facet combination. Some combinations had few coded references for one or both persona groups, so finding no difference reflects a lack of evidence rather than evidence of no inclusivity issue. 

\textit{External Validity:} Our study involved 20 student participants from a single university, which limits the extent to which findings can be generalised to a broader population or professional developers. As a qualitative study, our goal was to identify interaction patterns and understand inclusivity issues, rather than statistical generalisation. The study also took place in a controlled laboratory setting, which may not reflect how developers interact with AI coding assistants in real-world environments. To mitigate this, we designed tasks to resemble common development activities. However, Copilot may be used differently on larger, legacy codebases. In addition, we focus on GitHub Copilot as a representative AI coding assistant, and our findings reflect the version of Copilot available at the time of data collection. While interactions may differ across tools and environments, many of the observed patterns may stem from common characteristics of AI coding assistants, such as inline suggestions and low-friction acceptance. Future studies should examine these patterns in additional tools with more diverse participants to validate our findings.

\textit{Internal Validity:} Participants varied in prior experience with programming and Copilot. While all participants were students, these differences may have influenced confidence and interaction style independently of cognitive style. To reduce this effect, tasks were designed to require only basic programming skills. The 40-minute time limit and observational setting may also have encouraged faster, accept-first behaviour. As all participants worked under the same conditions, this cannot by itself explain differences between Abi-like and Tim-like participants, although it may have amplified them. Participants chose which Copilot features and modes to use, so some identified differences may reflect these choices. Finally, task order was fixed: code generation, debugging, then refactoring. As participants gained familiarity with Copilot throughout the session, later tasks may have been influenced by learning effects, reducing the discoverability of some inclusivity issues. 

\textit{Reliability:} 
Several members of the research team have prior experience applying the GenderMag framework in empirical software engineering research. This familiarity informed the study design and interpretation of the findings, but also created the possibility of confirmation bias. To mitigate this risk, we conducted the thematic analysis inductively before applying the GenderMag framework, allowing themes to emerge from the data rather than being constrained by pre-existing cognitive style categories. Qualitative analysis is inherently subjective and may be influenced by researcher bias. Coding and theme development involved multiple researchers and iterative reviews to challenge and refine interpretations.


%
\section{Conclusion}

This study investigated GitHub Copilot from a cognitive inclusion perspective, examining how developers with different cognitive styles experience and interact with the tool. Using the GenderMag framework, we identified five inclusivity issues. Our findings show that Copilot required additional effort for users who prefer understanding and certainty before acting, have lower self-efficacy, or want to learn from its suggestions. We proposed five design implications for more inclusive AI coding assistants: supporting access to reasoning, evaluation before acceptance, learning-oriented interactions, feature discoverability, and different levels of AI guidance and control. These design implications provide directions for developing more inclusive AI coding assistants.

\section{Data Availability}
More details on study procedures, including the questions in the expression of interest questionnaire, the pre-defined questions for the semi-structured interviews, and the outline for the lab sessions are in our online replication package~\cite{replication_arXiv}. We do not have participant consent to share transcripts.

\bibliographystyle{IEEEtran}
\addcontentsline{toc}{section}{\refname}\bibliography{RefDatabase}

\newpage

\end{document}